\documentclass[groupedaddress, superscriptaddress, showpacs,amsmath, amssymb, twocolumn,floatfix]{revtex4}  
\pdfoutput=1

\newif\ifpdf
  \ifx \pdfoutput\undefined 
    \pdffalse
  \else 
    \pdfoutput=1 \pdftrue 
  \fi
\ifpdf
  \usepackage[pdftex]{color,graphicx}
  \DeclareGraphicsExtensions{.pdf,.jpg,.png}
\else
  \usepackage{color,graphicx}
  \DeclareGraphicsExtensions{.ps,.eps}
\fi

\begin{document}

\title{Modifications of turbulence and turbulent transport associated
  with a bias-induced confinement transition in LAPD}
\author{T.A. Carter}
\email{tcarter@physics.ucla.edu}
\affiliation{Department of Physics and Astronomy, University of
  California, Los Angeles, CA 90095-1547}
\affiliation{Center for Multiscale Plasma Dynamics, University of California, Los Angeles, CA 90095-1547}
\author{J.E. Maggs}
\affiliation{Department of Physics and Astronomy, University of California, Los Angeles, CA 90095-1547}

\begin{abstract}

  Azimuthal flow is driven in the edge of the Large Plasma Device
  (LAPD) [W.~Gekelman, {\itshape et. al}, Rev. Sci. Instr. {\bfseries
      62}, 2875 (1991)] through biasing a section of the vacuum vessel
  relative to the plasma source cathode.  As the applied bias exceeds
  a threshold, a transition in radial particle confinement is
  observed, evidenced by a dramatic steepening in the density profile,
  similar to the L- to H-mode transition in toroidal confinement
  devices.  The threshold behavior and dynamic behavior of radial
  transport is related to flow penetration and the degree of spatial
  overlap between the flow shear and density gradient profiles.  An
  investigation of the changes in turbulence and turbulent particle
  transport associated with the confinement transition is presented.
  Two-dimensional cross-correlation measurements show that the spatial
  coherence of edge turbulence in LAPD changes significantly with
  biasing. The azimuthal correlation in the turbulence increases
  dramatically, while the radial correlation length is little altered.
  Turbulent amplitude is reduced at the transition, particularly in
  electric field fluctuations, but the dominant change observed is in
  the cross-phase between density and electric field fluctuations.
  The changes in cross-phase lead to a suppression and then apparent
  reversal of turbulent particle flux as the threshold is exceeded.

\end{abstract}


\maketitle

\section{Introduction\label{intro}}

Turbulence has long been recognized as the dominant cause of
cross-field transport in magnetically-confined
plasmas~\cite{Liewer:1985} Significant progress in the effort to
confine plasmas for fusion energy came with the discovery of the
high-confinement mode or ``H-mode,'' where the spontaneous formation
of an edge transport barrier occurs after the heating power exceeds a
threshold~\cite{Wagner:1982}.  A key characteristic of these
transport barriers is the presence of a localized, non-uniform radial
electric field with associated cross-field $E\times B$ flow and flow
shear~\cite{Burrell:1997, Terry:2000}.  
The important role of
radial electric field in the confinement transition was first
demonstrated clearly in experiments in the Continuous Current Tokamak
(CCT)~\cite{Taylor:1989,Tynan:1996} where biasing was used to
establish a radial electric field and trigger an H-mode confinement
transition.  Since these pioneering experiments, biasing has been used
on a number of confinement devices to study transport barrier
formation and turbulence
modification~\cite{Weynants:1992,Boedo:2000,Silva:2006,Craig:1997,Chapman:1998,Shats:2000,Sakai:1993}.

Flow and flow shear can modify turbulence and
transport through a number of processes.  First, the shear can lead to
radial decorrelation or ``shearing apart'' of turbulent eddies, reducing
their radial extent and therefore transport effectiveness~\cite{Biglari:1990}.
Also, the amplitude of turbulent fluctuations can be reduced
through the modification of the nonlinearly saturated level of the
turbulence~\cite{Burrell:1997}.  Finally, electrostatically-driven particle
transport is dependent on the correlation between density and electric
field fluctuations in low frequency turbulence (where $E\times B$ is
the dominant particle response):
\begin{equation}
\begin{split}
\Gamma_n & = \left<n v_{\rm r}\right> = \frac{\left<n
    E_\theta\right>}{B} \\ 
& = \frac{2}{B}\int_0^\infty |n(\omega)||{E_\theta}(\omega)|
    \gamma_{n,E_\theta}(\omega)
    \cos\left(\theta_{n,E_{\theta}}(\omega)\right) d\omega
\end{split}
\label{eqn1}
\end{equation}

where $n(\omega)$ and $E_\theta(\omega)$ are the Fourier transforms of
density and azimuthal electric field, $\gamma_{n, E_\theta}$ is the
cross-coherency between $n$ and $E_\theta$, and $\theta_{n,E_\theta}$
is the cross-phase between density and electric field
fluctuations~\cite{Powers:1974}. Flow and flow shear can modify both
the cross-phase and cross-coherency between density and potential
fluctuations, leading to a reduction of transport even in the absence
of amplitude reduction~\cite{Ware:1996,Moyer:1995}.

Changes in turbulence and transport associated with azimuthal flow
have been investigated in the Large Plasma Device (LAPD) at UCLA.
Flows are driven through biasing the vacuum chamber wall relative to
the plasma source~\cite{Maggs:2007,Horton:2005,Perez:2006}.
Excitation of the Kelvin-Helmholtz instability has been observed in
LAPD when a strong flow gradient is imposed (through modification of
the magnetic topology)~\cite{Horton:2005,Perez:2006}.  For flow
driven in a straight magnetic field line configuration in LAPD
(resulting in smaller but non-negligible flow shear), a transition in
particle confinement is observed.  Models of diffusive particle
transport and changes in transport are presented in an earlier
publication~\cite{Maggs:2007} where it is reported that transport
rates change from Bohm to classical due to edge rotation. The
observation of a confinement transition in a simple magnetic geometry
provides the opportunity to perform a detailed study of modifications
of turbulence and turbulent transport in a system free from
complications associated with toroidal systems (e.g. field line
curvature, poloidal asymmetry, trapped particles).  Good diagnostic
access to LAPD provides for detailed measurements of the
spatial and temporal characteristics of the turbulence.

Here we summarize the primary results reported in this paper.
Measured turbulent transport flux is reduced and then suppressed,
leading to a confinement transition, as the applied bias is increased.
The threshold in the applied bias is linked to radial penetration of
the driven azimuthal flow.  Two-dimensional measurements of the
turbulent correlation function show that the azimuthal correlation
increases dramatically during biasing, with the high m-number modes
involved in turbulent transport becoming spatially coherent.  However
no significant change in the radial correlation length is observed
associated with the confinement transition.  As the bias is increased
above threshold, there is an apparent reversal in the particle flux
(indicating inward transport). The peak amplitude of density and
electric field fluctuations do decrease, but the reduction is only
slight and does not fully explain the transport flux reduction. The
cross-phase between density and electric field fluctuations changes
significantly as the threshold for confinement transition is reached,
and explains the reduction and reversal of the measured transport
flux.  The dynamics of the transition have also been studied and show
a correlation between transport suppression and the overlap of the
flow (shear) profile and the density gradient profile.

\section{Experimental Setup\label{setup}}

\begin{figure}[!htbp]
\includegraphics[width=3.0in]{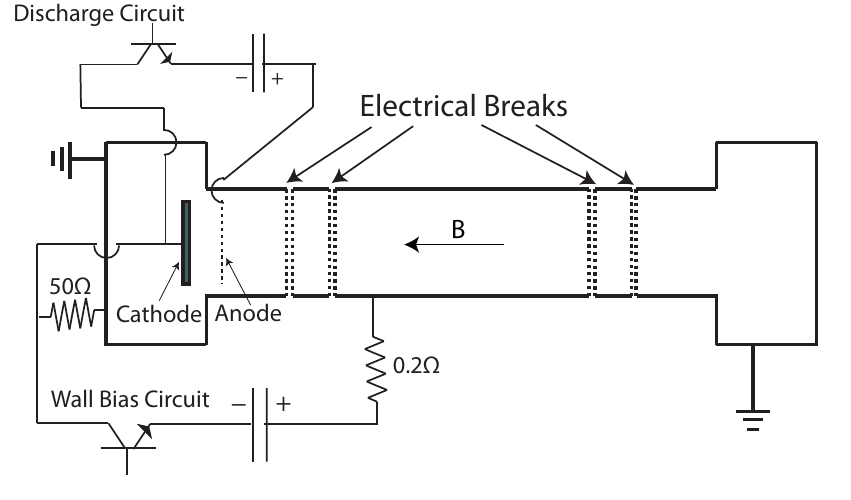}
\caption{Schematic of LAPD including wall biasing
  circuit.}
\label{fig1}
\end{figure}

The experiments were performed in the upgraded Large Plasma Device
(LAPD)~\cite{Gekelman:1991}, which is part of the Basic Plasma Science Facility
(BaPSF) at UCLA. The vacuum chamber of LAPD is 18~m long and 1~m in
diameter and surrounded by solenoidal magnetic field coils.  The
plasma is generated by a cathode discharge~\cite{Leneman:2006}.  The
cathode is 73~cm in diameter and is located at one end of the vacuum
chamber, as shown schematically in Fig.~\ref{fig1}.  A molybdenum
mesh anode is situated 50~cm away from the cathode.  A bias of 40-60V
is applied between the cathode and anode using a solid-state
switch~\cite{Pribyl:2004}, resulting in 3-6kA of discharge current.  An
important aspect of the operation of the plasma source is the
generation of primary electrons with energy comparable to the
anode-cathode bias voltage.  The mesh anode is 50\% transparent,
allowing half of these primaries to travel down the magnetic field
into the main chamber leading to ionization and heating of the bulk
plasma.  The cathode is situated at the mouth of the solenoid and
therefore sits in a region of flaring magnetic field.  The flared
magnetic field maps the 73~cm diameter cathode to a $\sim 56$~cm
diameter region in the main chamber.  Primary electrons from the
source are isolated to this region in the chamber and lead to a fast
electron tail for $r \lesssim 28$cm.  Typical plasma
parameters in LAPD discharges are $n_e \lesssim 5 \times 10^{12}$
cm$^{-3}$, $T_e \sim 7$~eV, $T_i \sim 1$~eV, and $B < 2$kG. In the
experiments reported here, the primary plasma species was
singly-ionized helium and the magnetic field strength used was 400G.
Modeling~\cite{Maggs:2007} and spectroscopic measurements show that the
ionization fraction of the LAPD plasma is $\gtrsim 50\%$ and therefore
Coulomb collisions are the most important collisional process.
However, neutral collisions (charge exchange in particular) are
important for establishing the radial current during biasing.

Measurements of density, temperature, floating potential and their
fluctuations are made using Langmuir probes.  A four-tip probe
with 0.76mm diameter tantalum tips arranged in a diamond pattern is
used as a triple Langmuir probe and particle flux probe. Two
tips are separated 3mm along the field and are used as a double probe
to measure ion saturation current ($I_{\rm sat} \propto n_e
\sqrt{T_e}$).  The remaining two tips are separated 3mm perpendicular
to the field and measure floating potential for deriving azimuthal
electric field fluctuations and temperature using the triple Langmuir
probe method~\cite{Chen:1965}.  Radial particle transport can be evaluated directly
using measured density and electric field fluctuations through
Eqn.~\ref{eqn1}. Flows are measured using a Gundestrup (Mach) probe with
six faces~\cite{Gunn:2001}.  The flow measurements are corrected for the
finite acceptance angle of the probe faces, which are smaller than the
ion gyroradius~\cite{Shikama:2005}.

 Using the difference in floating potential to determine the azimuthal
electric field is problematic in the presence of fast electron tails,
such as exist on field lines connected to the cathode source.
In the presence of fast electrons,
differences in floating potential may no
longer be proportional to differences in plasma
potential and thus may not accurately measure the electric field.
In addition, the fast electron tail, which sets floating
potential, is much less collisional than the bulk plasma so that the
floating potential is no longer a localized quantity. This
de-localization skews the correlation with locally measured density
fluctuations and thereby affects the
flux measurement.
Measurements of $I_{\rm sat}$ are made with a double-Langmuir probe biased to
70V to reject primary electrons, and therefore should not be affected
by the fast electron tail. Thus, in this study, we report on the
properties of ion saturation fluctuations everywhere in the plasma
column. Properties of electric field fluctuations and
cross-correlation flux measurements are presented everywhere in the
plasma column, but a caution is issued in regard to measurements made
on field lines where primary electrons are present ($r \lesssim 28$cm).

\begin{figure}[!htpb]
\includegraphics[width=3.0in]{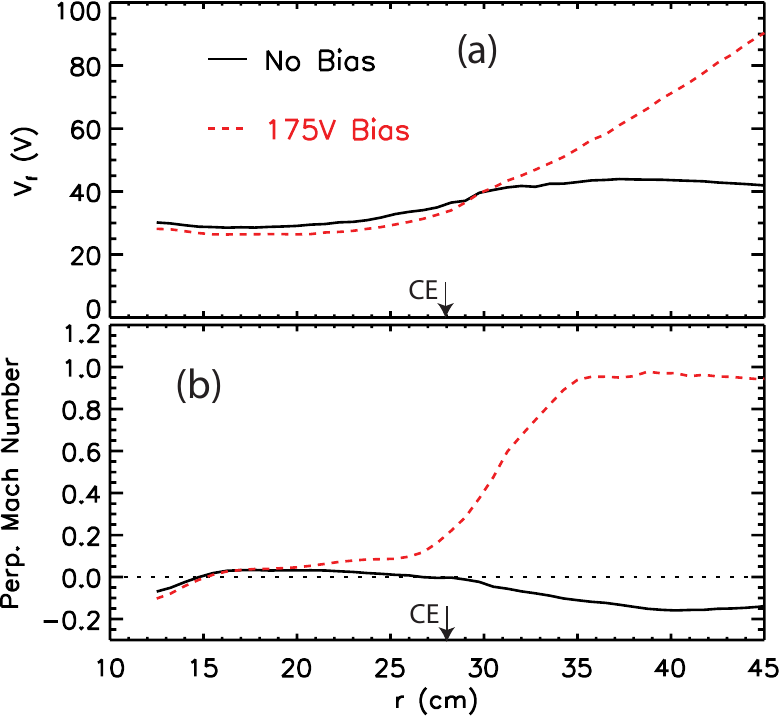}
\caption{(color online) Radial profiles of {\bfseries (a)} floating
  potential and  {\bfseries (b)} perpendicular Mach number for no bias
  and 175V bias.  Note the radial location of field lines that connect
to the cathode edge are marked with ``CE'' in this figure and
subsequent figures.}
\label{fig2}
\end{figure}

The edge plasma in LAPD is rotated through biasing the vacuum vessel
wall positively with respect to the source cathode. Figure~\ref{fig1}
shows a schematic of the LAPD along with the rotation bias circuit,
which includes a capacitor bank and an IGBT (Insulated Gate Bipolar
Transistor) switch.  The rotation bias is pulsed, with the switch
closed for 4~ms during the steady portion (current flattop) of the
LAPD discharge.  Typical bias voltages are on the order of 100V, and
result in around 100A of peak radial current.  The radial current
provides the torque (through $J\times B$ forces) needed to spin up the
plasma. The measured current is consistent with ions carrying the
radial current through ion-neutral collisions (Pederson
conductivity)~\cite{Maggs:2007}.  The ions carry the current radially
until they reach field lines which connect to the cathode in the
plasma source, where the current closes via parallel electron
currents.  The influence of the applied wall bias is therefore limited
to regions of the plasma that are not connected along the magnetic
field to the cathode source.  This is demonstrated in
Figure~\ref{fig2}~(a), which shows the measured floating potential as
a function of radius for no bias and a bias of 175V.  Measurements
shown in this figure (and subsequent figures) are made with a 1~cm
radial resolution through moving a single probe on a shot-to-shot
basis.  The trace shown is the result of a time average over 1.3~ms of
data during the flat-top of the LAPD discharge current pulse. Field
lines in the region $r\lesssim 28$~cm connect to the cathode, and the
floating potential is unchanged by the applied bias in this region.
Note that in this figure (and many following figures) the radial
location of the field lines which connect to the cathode edge is
marked with ``CE.''  Measurements of azimuthal flow (Mach number) for
the same case are shown in Fig~\ref{fig2}~(b), indicating that driven
flows are also localized to field lines not connected to the plasma
source.

\section{Experimental Results\label{results}}
\subsection{Threshold for confinement transition\label{threshold}}

\begin{figure}[!htbp]
\includegraphics[width=3.0in]{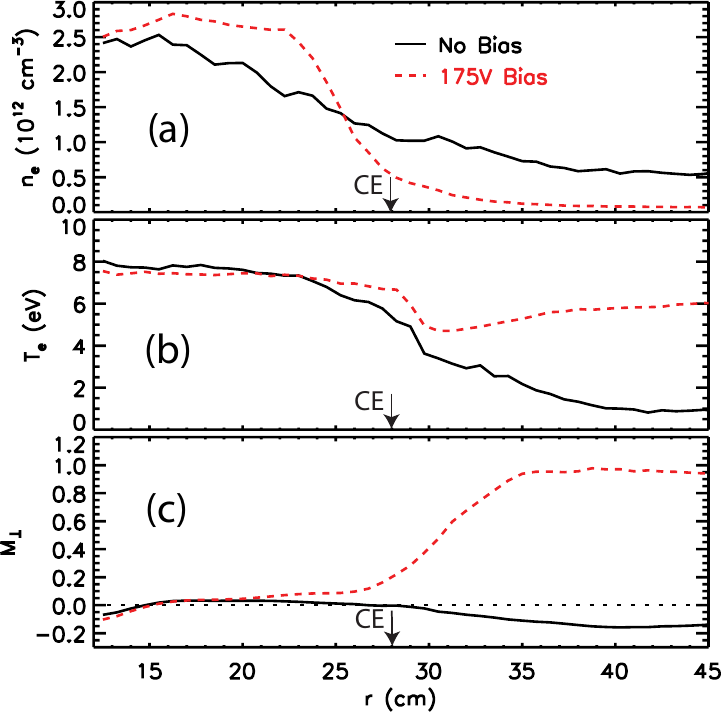}
\caption{(color online) Profile of {\bfseries(a)} density, {\bfseries
    (b)}  Electron temperature, and (c) Mach number before and
  after a confinement transition.}
\label{fig3}
\end{figure}

A radial particle confinement transition is observed with
biasing~\cite{Maggs:2007}.  Figure~\ref{fig3} shows profiles of plasma
density, ion saturation current ($I_{\rm sat}$) fluctuations, and
perpendicular Mach number for two cases: unbiased and 175V
bias. Without bias, radial turbulent particle transport causes the
density profile to extend well past the cathode edge, with a fairly
gentle gradient ($L_n = |\nabla \ln n|^{-1} \sim 10$~cm), as shown in
Fig.~\ref{fig3}~(a). For comparison, the ion sound gyroradius is
$\rho_s \sim 1.2$~cm (helium, $T_e \sim 6$eV, $B=400$G) and the ion
gyroradius is $\rho_i \sim 0.5$~cm ($T_i \sim 1$eV).  In the biased
case, the measured density profile steepens dramatically ($L_n \sim
4$~cm). Detailed transport modeling of the LAPD plasma shows that the
profile before biasing is consistent with radial diffusion at the Bohm
rate.  The steepened profile after biasing is consistent with
classical collisional diffusion, a factor of $\sim 100$ drop in the
diffusion coefficient~\cite{Maggs:2007}. In the rotating plasma, the
density profile is essentially set by the plasma production source
profile since the radial transport rates are so slow.

 The electron temperature profiles measured using the triple probe
 technique are shown in Fig. 3 (b). The electron temperature drops
 rapidly past the edge of the cathode. The electron beam from the
 cathode acts primarily as a heat source once the plasma is formed.
 The fully formed plasma column is hot and dense enough that the tail
 of the Maxwellian distribution dominates the beam electrons in the
 ionization process. In the rotating plasma the electron temperature
 is elevated in the region outside the plasma source, perhaps due to
 frictional heating from neutral drag. The Mach-probe measured
 azimuthal flow profile, with and without bias, is repeated in
 Fig.~\ref{fig3}~(c) for reference.

%\begin{wrapfigure}{l}{2.5in}

\begin{figure}[!htbp]
\includegraphics[width=3.0in]{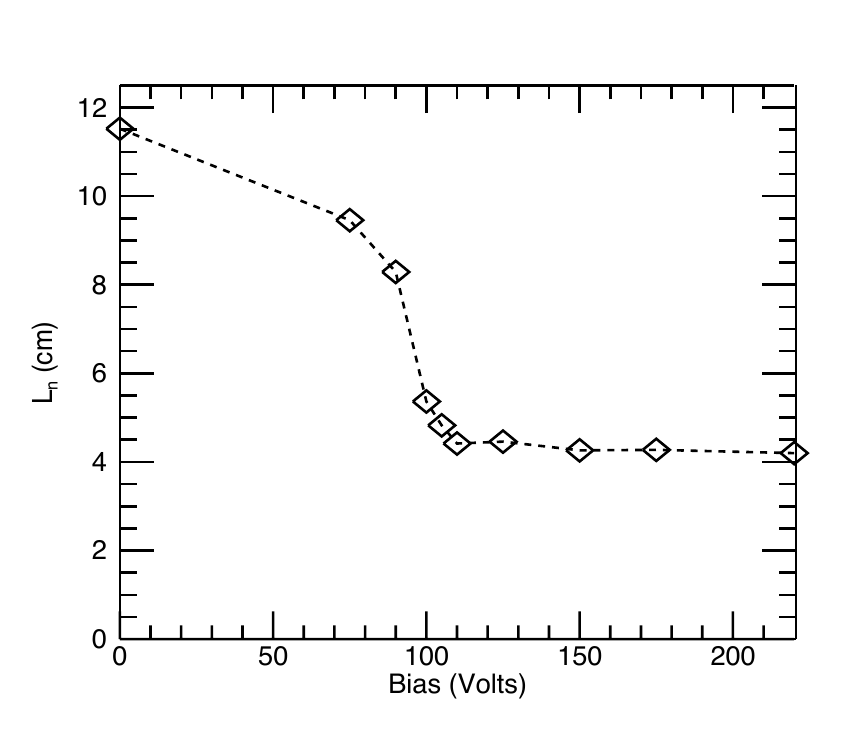}
\caption{Minimum radial density gradient scale length as a function of
applied bias. (This figure reprinted from reference~\cite{Maggs:2007})}
\label{fig4}
\end{figure}

The observed profile steepening occurs for bias voltages above a
threshold value.  Figure~\ref{fig4} shows the minimum density
gradient scale length ($L_n$) versus bias voltage.  For biases above
$\sim 100$V, the density gradient scale length decreases dramatically,
indicating a transition in the radial particle confinement.  As the
bias is increased, the density gradient scale length saturates (near
$L_n \sim 4$~cm), consistent with modeling which shows that transport
beyond the confinement transition threshold is suppressed to the
irreducible lowest level set by classical transport~\cite{Maggs:2007}.  

\begin{figure}[!htbp]
\includegraphics[width=3.0in]{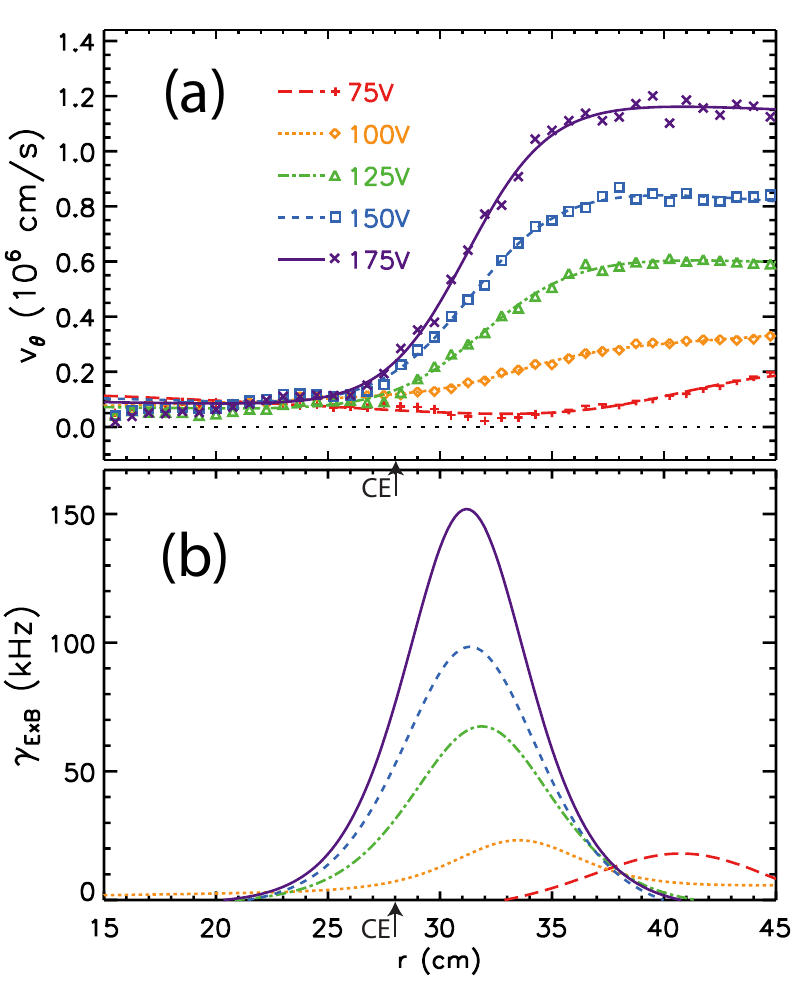}
\caption{(color online) For several bias values: (a) Radial profile of azimuthal velocity, with fit. (b)
  Radial profile of shearing rate, derived from fit to azimuthal
  velocity profile.}
\label{fig5}
\end{figure}

The threshold for confinement transition in LAPD appears to be set by
radial penetration of the bias-driven flow.  Figure~\ref{fig5}(a)
shows the radial profile of azimuthal flow for several values of the
bias, including a fit to the data of the form $a_1\tanh((r - a_2)/a_3)
+ a_4r^2 + a_5$.  Figure~\ref{fig5}(b) shows the profile of the
shearing rate, $\gamma = \partial v_\theta/\partial r$, derived from
the fit.  This technique is used to reduce the noise introduced by
taking radial derivatives of the raw data. Below the transition
threshold (75V) the flow and flow shear is concentrated in the far
edge plasma, and has not penetrated in to the cathode edge at $r\sim
28$cm.  As the threshold is reached (100V) the peak value of the flow
shear has not increased beyond the below-threshold value, but the flow
has moved radially inward so that some finite flow shear exists near
the cathode (plasma source) edge.  
As the bias is further increased, the flow
amplitude and shear increase, but the shape of the profile is
relatively unchanged.  The observed transition is therefore not due to
systematic increase of the shearing rate through a threshold value,
but is instead associated with a sudden appearance of a large shearing
rate at the plasma source edge.

\begin{figure}[!htbp]
\includegraphics[width=3.4in]{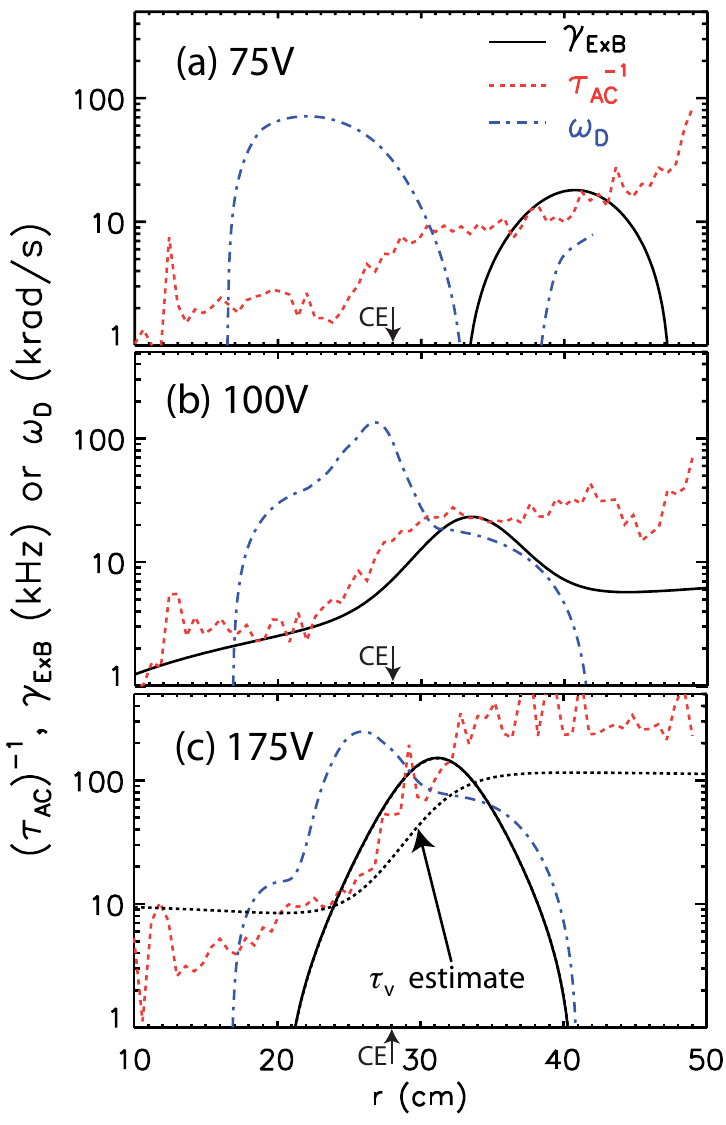}
\caption{(color online) Radial profiles of the shearing rate
  ($\gamma_{\rm E\times B}$), the turbulent autocorrelation time ($\tau_{\rm ac}$), and
  the drift frequency ($\omega_D$) for three bias values:  75V (below threshold), 100V
  (threshold) and 175V (above threshold).}
\label{fig6}
\end{figure}

 If the change in radial transport rates is associated with shear
 suppression of turbulent transport, the shearing rate should be large
 enough to compete with characteristic timescales of the turbulence,
 often taken to be the linear growth rate or the nonlinear eddy
 turnover time.  The turbulence in the edge of LAPD is known to be
 drift-Alfv\'{e}n in nature~\cite{Maggs:2003}, and the characteristic
 frequency of drift waves is the diamagnetic drift frequency,
 $\omega_{D} = k_y v_{\rm th,e} \rho_{\rm e}/L_n$ (where $k_y$ is the
 azimuthal wavenumber, $v_{\rm th,e}$ is the electron thermal speed
 and $\rho_{\rm e}$ is the electron gyroradius).  The drift frequency
 will be taken as a proxy for the linear growth rate, although the
 growth rate of resistively-driven modes is often only a small
 fraction of this frequency~\cite{Horton:1999}.  The nonlinear
 timescale in the turbulence, the eddy turnover time, can be estimated
 from measurements of the autocorrelation time in the turbulence,
 ${\tau_{\rm ac}}$. Figure~\ref{fig6} shows the radial profile of
 shearing rate, $\tau_{\rm ac}^{-1}$ and $\omega_{D}$ for three bias
 values: 75V (below threshold), 100V (near threshold) and 175V (above
 threshold). In the calculation of {${\omega_D}$}, the effective
 azimuthal wave number of the fluctuations, $k_y$, is taken from
 spatial correlation measurements (detailed later in this paper) which
 show a dominant azimuthal mode number of $m\sim 10$. The
 autocorrelation time is derived from the lab-frame autocorrelation
 function computed from ion saturation current data, which is fit to
 the function $a_0 \cos\left(a_1\tau\right) \exp\left(-a_2 \left|
 \tau\right|\right)$, where $\tau_{\rm ac} = 1/a_2$.  At the threshold
 for transition, the shearing rate in the edge region is comparable to
 the local diamagnetic drift frequency and the autocorrelation
 frequency (1/${\tau_{\rm ac}}$).  In a flowing plasma, the
 autocorrelation time in the lab frame is affected both by flow of
 turbulent structures past the probe as well as temporal decorrelation
 of turbulent structures in the plasma frame (the eddy
 lifetime)~\cite{Bencze:2005}.  The autocorrelation time due to flow
 of structures past the probe can be estimated as $\tau_v \sim
 w_\perp/v_{\rm flow}$, where $w_\perp$ is the azimuthal scale size of
 the turbulence (azimuthal correlation length) and $v_{\rm flow}$ is
 the azimuthal flow velocity.  An estimate of $\tau_{\rm v}^{-1}$ is
 shown for a bias of 175V in Fig.~\ref{fig6}(c), using an azimuthal scale size
 of $w_\perp \sim 10$cm (motivated by spatial correlation
 measurements). The $\tau_{\rm v}^{-1}$ estimate is comparable to the
 calculated $\tau_{\rm ac}^{-1}$, indicating that flow of structures
 past the probe is a major contributer to establishing the lab-frame
 autocorrelation time, and that it can not be interpreted as a measure
 of the eddy turnover time.  However, $\tau_{\rm ac}$ does set a lower
 limit for the eddy turnover time and the actual plasma-frame eddy
 lifetime must be longer than $\tau_{\rm ac}$ (and therefore
 $\tau_{\rm eddy}^{-1}$ must be smaller than $\tau_{\rm ac}^{-1}$).
 It is therefore reasonable to argue that the flow shear is large
 enough to be dynamically important and thus could suppress turbulent
 transport, leading to the observed confinement transition.

\subsection{Changes to spectrum and transport flux under biasing\label{spec}}

\begin{figure}[!htbp]
\includegraphics[width=3.4in]{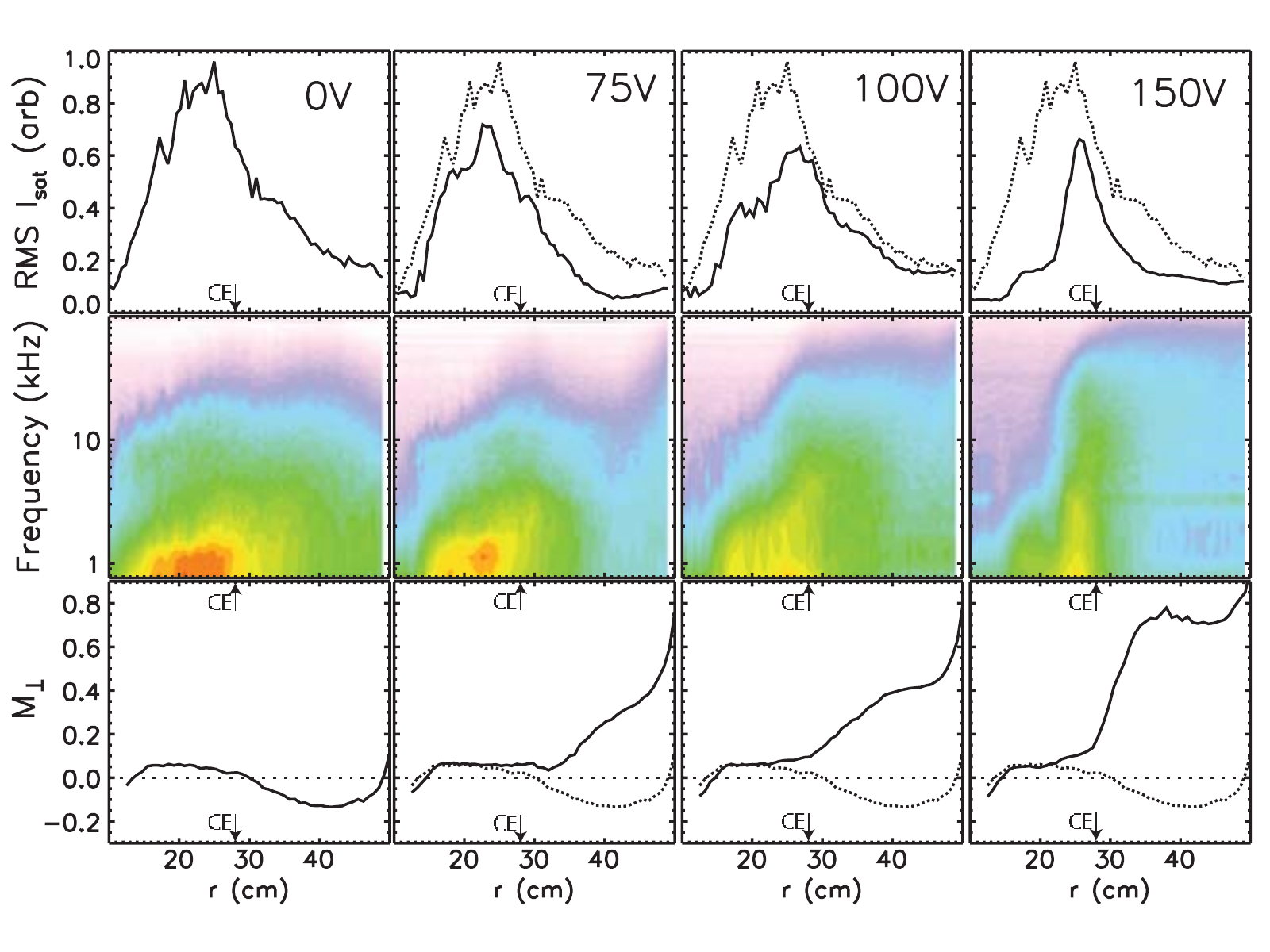}
\caption{(color online) Radial profile of RMS $I_{\rm sat}$ fluctuations, Contours of $I_{\rm sat}$ power
spectrum versus radius and frequency, and Mach number profile for
several bias values. The radial profiles of RMS fluctuation level and
Mach number in the unbiased case are shown as dashed lines for
comparison with the biased cases.}
\label{fig7}
\end{figure}

Detailed measurements of turbulent spectra and turbulent particle flux
have been performed to evaluate changes associated with the
confinement transition. Figure~\ref{fig7} shows the radial profile of
rms $I_{\rm sat}$ fluctuations, contours of the $I_{\rm sat}$ power
spectrum, and the profile of perpendicular Mach number for several
values of the rotation bias. All are time-averaged over the length of
the bias pulse. The normalized fluctuation amplitude is $\delta I_{\rm
  sat}/I_{\rm sat} \sim 33\%$ at the peak of the fluctuation amplitude
in the unbiased case.  The fluctuation spectrum is broadband prior to
biasing, with no evidence for coherent modes. Below threshold (75V),
some reduction of fluctuation amplitude is seen in the far edge, as
well as a Doppler upshift in the fluctuation spectrum (especially
associated with the strong flow feature near the wall at $r =
50$cm). As the transition threshold is approached (100V) and exceeded
(150V), the fluctuation amplitude is reduced and concentrates on the
steepened density gradient, but the spectrum remains broadband. The
rearrangement of the fluctuation profile accompanies changes in the
plasma density profile (not shown). A strong Doppler shift is also
observed in the spectrum localized to the region of strong azimuthal
flow.

\begin{figure}[!htbp]
\includegraphics[width=3.4in]{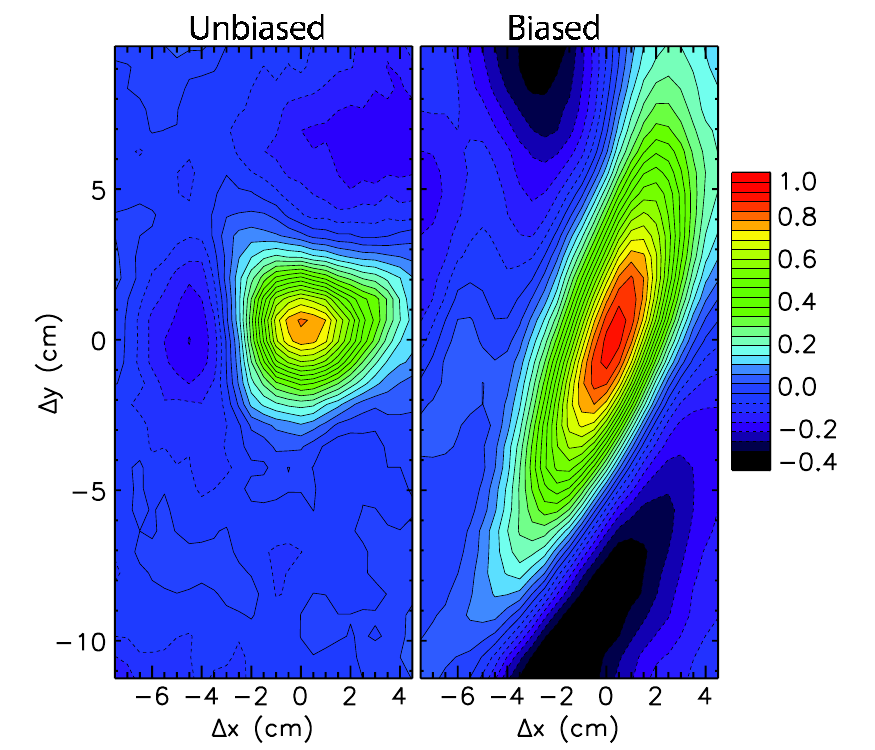}
\caption{(color) Two-dimensional cross-field correlation function for $I_{\rm
    sat}$ fluctuations in unbiased and biased plasmas}
\label{fig8}
\end{figure}

Radial decorrelation, or reduction of the radial size of turbulent
eddies, is one possible mechanism by which turbulent transport might
be suppressed by sheared flow. Measurements of the two-dimensional
(cross-field) turbulent correlation function have been performed to
evaluate the role of this mechanism in the confinement
transition. Figure~\ref{fig8} shows measurements of the
two-dimensional correlation function of fluctuations in ion saturation
current for unbiased and strongly biased (220V) cases. These
measurements are made using two probes separated along the magnetic
field by 0.69m. One probe is fixed in space near the cathode edge (at
$r = 26$cm) and used as a reference to correlate against, while the
second probe is moved shot-to-shot to make measurements at 775
locations in the plane perpendicular to the background magnetic field
(the density gradient is in the $-\hat{x}$ direction). Prior to
biasing, the correlation function is nearly isotropic, with a slightly
longer correlation length in the radial direction ($\hat{x}$ in the
figure). The correlation lengths in the unbiased case (radial and
azimuthal) are $\lesssim 5$~cm, around 5 times the ion sound gyroradius
and 10 times the ion gyroradius.  Under biasing, the azimuthal
($\hat{y}$) correlation length increases dramatically. However, the
radial extent of the correlation function is not reduced significantly
and therefore no strong evidence for radial decorrelation is found. In
addition, a clear pattern of maxima and minima are visible in the
correlation pattern measured in the rotating plasma. This pattern
resembles the pattern expected of a coherent, high m-number
cylindrical eigenmode.

 There are two potential explanations for transport reduction based on
 the measured correlation function. First, because the correlation
 function is strongly elongated in the presence of the flow, the eddy
 turnover time is lengthened: it takes longer for a blob of plasma to
 travel from one side of the eddy to the other in the radial
 direction.  So the effective transport time step is lengthened
 ({\itshape i.e.} the eddy turnover time increases), even though the
 step size (the radial correlation length) is largely unchanged.  This
 explanation seems inconsistent with Fig.~\ref{fig6}, which shows the
 autocorrelation time decreasing with increasing bias.  However, the
 lab-frame autocorrelation time may be dominated by flow effects.
 Second, the spatial coherency of the fluctuations has increased
 dramatically as a result of the biasing. Turbulent particle diffusion
 is effective only in the presence of decorrelation: coherent vortices
 can flatten the density profile over their width through mixing, but
 do not lead to radial diffusion.

\begin{figure}[!htbp]
\centerline{
\includegraphics[width=3.4truein]{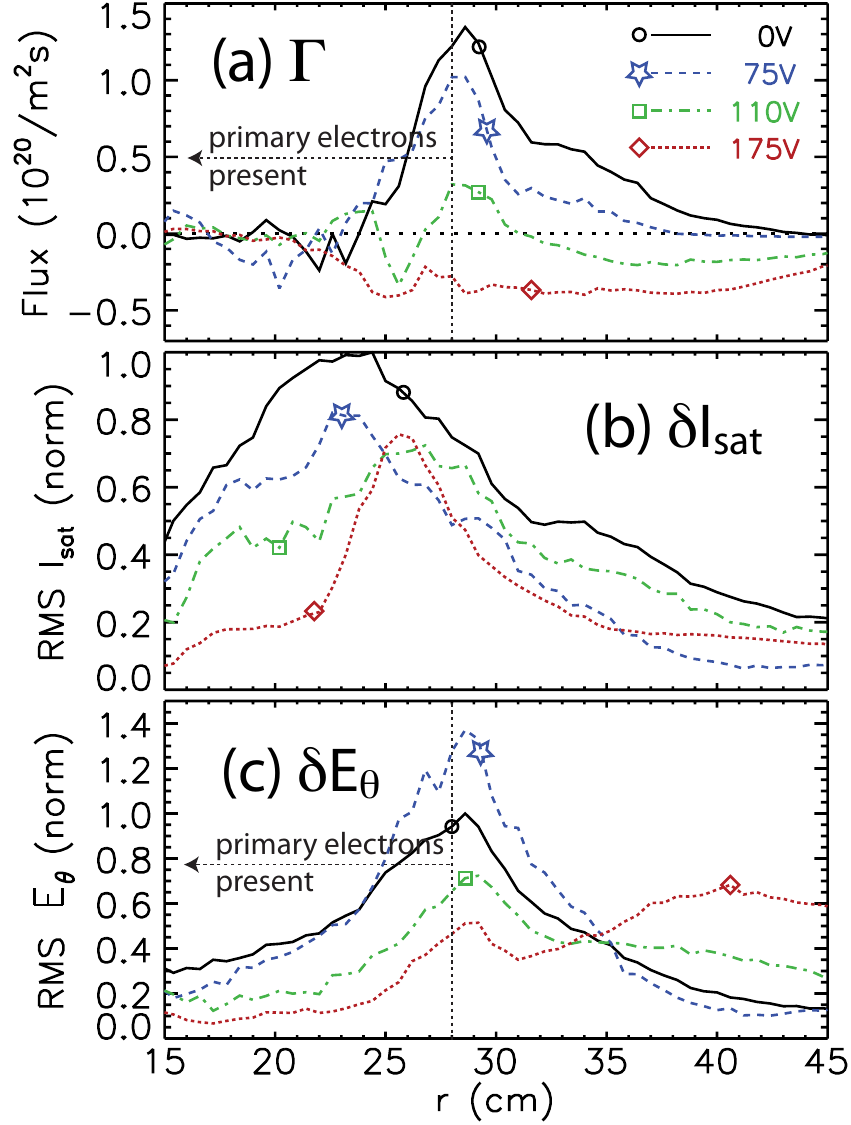}}
\caption{(color online) (a) Particle flux profile, (b)
root-mean-square (RMS) $I_{\rm sat}$ fluctuation
profile, and (c) RMS $E_\theta$ fluctuation profile for several bias
values.}
\label{fig9}
\end{figure}

The expression for particle flux given in Eq.~\ref{eqn1} can be
directly evaluated using data obtained by a particle flux probe. Ion
saturation current is measured to determine density fluctuations while
floating potential is measured at two azimuthally separated probe tips
in order to estimate fluctuations in the azimuthal electric field
$E_\theta$. Figure~\ref{fig9} (a) shows the radial profile of
turbulent particle flux measured for four different bias voltages. The
flux is computed using the FFT of ion saturation current and electric
field fluctuations, taking the final 2.6~ms of the 4~ms bias pulse as
the window for the FFT computation.  The measured particle flux is
reduced and then suppressed as the bias is increased to 110V (just
past the threshold voltage). As the bias is further increased, there
is a reversal in the measured particle flux indicating inward
transport. The particle flux in Eq.~\ref{eqn1} is established by the
fluctuation amplitudes and the cross-coherency and cross-phase between
density and electric field fluctuations. Figure~\ref{fig9} (b) and (c)
show the radial profiles of density and electric field fluctuations
for the same bias voltages. The density fluctuations become more
localized in space (peaked on the steepened density gradient), but the
peak amplitude does not decrease significantly. The electric field
fluctuations in the gradient region are also slightly reduced after
the threshold is exceeded. However, electric field fluctuations grow
stronger in the far edge with increased bias.  As noted earlier,
measurements of electric field fluctuations (and hence computed
particle flux) are suspect in the presence of primary electrons ($r
\lesssim 28$cm).  The radial region in which primary electrons are
present is specifically marked in Fig.~\ref{fig9} (a) and (c).

\begin{figure}[!htbp]
\centerline{
\includegraphics[width=3.4truein]{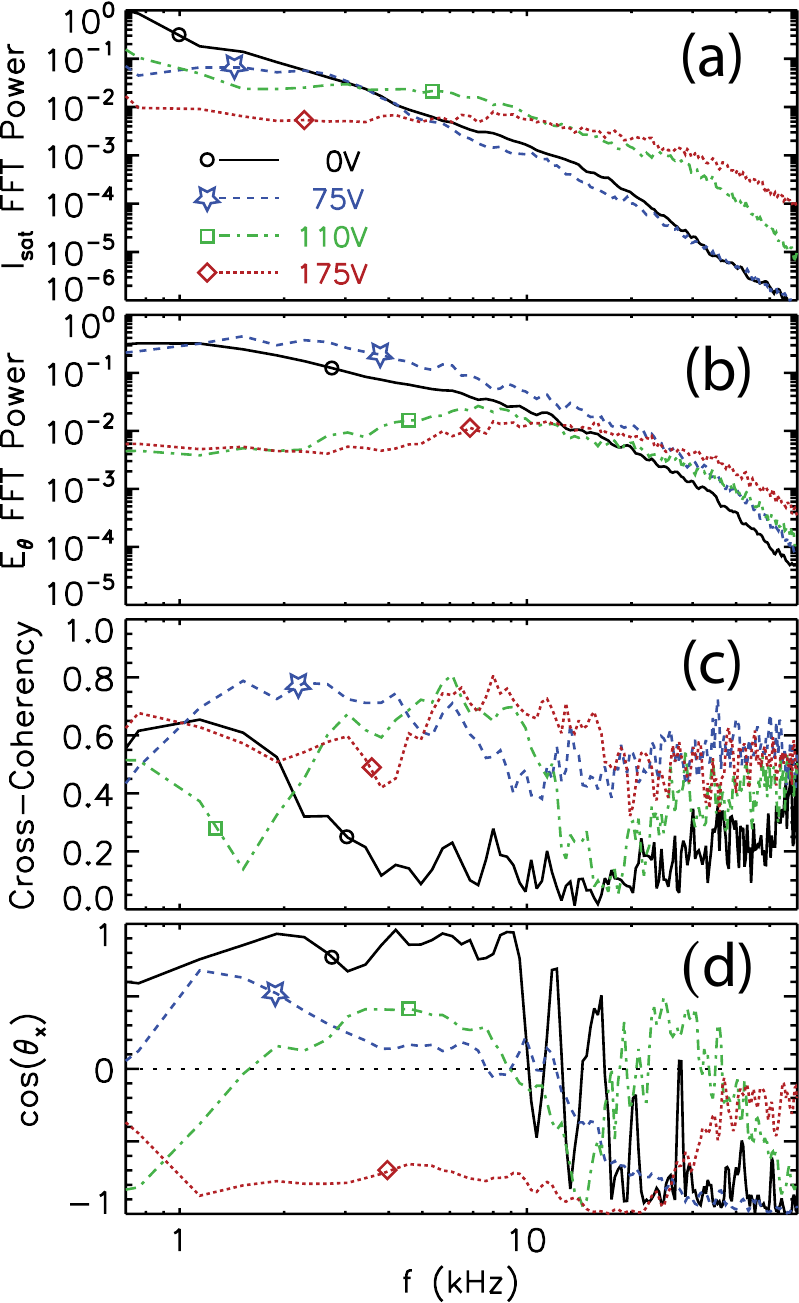}}
\caption{(color online) For the same bias values, averaged over $28 <
  r < 30$cm and using a 2.6~ms time window at the end of the bias
  pulse for the FFT: (a) $I_{\rm sat}$ FFT power spectrum, (b)
  $E_\theta$ FFT power spectrum, (c) Cross-coherency and (d)
  cross-phase between $I_{\rm sat}$ and $E_\theta$. }
\label{fig10}
\end{figure}

The fast Fourier transform (FFT) power spectrum of $I_{\rm sat}$
fluctuations is shown in Fig.~\ref{fig10}(a)
(averaged over spatial locations 28cm {\textless} r {\textless} 30cm).
 The spectrum is broadband before and after the transition, with
decreased power at low frequency and a Doppler upshift in the spectrum
apparent with increasing rotation. The cross-coherency between density
and electric field fluctuations actually increases with bias,
as shown in Fig.~\ref{fig10} (c). Although the electric field
fluctuation amplitude does decrease, the dominant cause of the
observed flux behavior with bias appears to be due to the
cross-phase. As shown in Fig.~\ref{fig10} (d), the cosine of the cross-phase
is very favorable for outward transport below the threshold, is
modified to near zero around the threshold, and finally
reverses sign at larger bias values.

\begin{figure}[!htbp]
\centerline{
\includegraphics[width=3.4in]{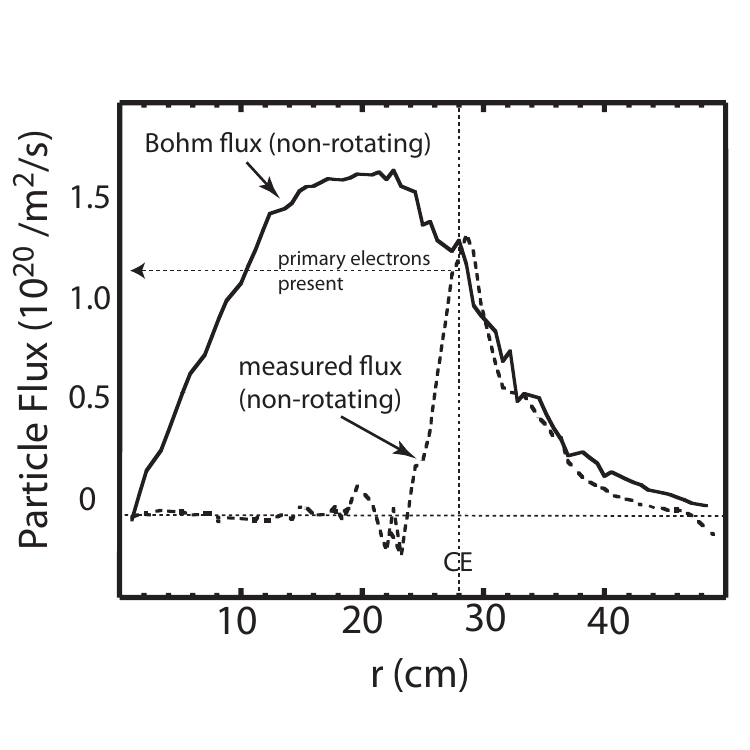} }
\caption{Comparison between model and measured fluxes.  Primary
  electrons may influence the ability to measure transport flux, and
  the region where they exist is marked in the figure.}
\label{fig11}
\end{figure}

Figure~\ref{fig11} shows a comparison between the radial
profiles of the measured flux and the flux determined from transport
  analysis~\cite{Maggs:2007}. In the unbiased, non-rotating case, the measured and
  predicted flux agree remarkably well on field lines that do not
  connect to the cathode ($r > 28$cm).  However, on cathode connected
  field lines (where primary electrons exist), the measured flux is
  near zero. One possible explanation for these observations is the
  influence of 50eV primary electrons from the source on the floating
potential measurement. The fastest electrons establish floating
potential, and on the cathode-connected field lines these are the
primary electrons and not the bulk electrons ($T_e \sim 5$eV). The primaries may not participate in the
drift-wave dynamics or if they do, they have extremely long mean free
path (comparable to the machine size) and any perturbation imparted on
them might be phase-mixed away as they are highly nonlocal in the wave.
Figure~\ref{fig9} is consistent with this argument, showing that the electric
field fluctuations peak outside of the cathode radius while
fluctuations in $I_{\rm sat}$ (which is measured
with $V_{\rm bias} \sim 70$V to reject primary electrons) peak further inward.

Inward flux (up the density gradient) is measured in these experiments
at larger values of wall bias, while the transport model flux of the
rotating plasma is nearly zero (classical transport).  Inward flux has
been reported in other biasing experiments on toroidal confinement
devices~\cite{Boedo:2002,Shats:2000} and also in the presence of
turbulently driven large-scale shear flows in a cylindrical
device~\cite{Tynan:2006}.  In addition, turbulence-driven particle
pinches are predicted theoretically in the presence of strongly
sheared flow~\cite{Terry:2003}.  However, if the turbulence is driven
purely by density gradients, only local reversal of the flux is
allowed, and not reversal across the whole profile, as observed here
(thermodynamically, the density gradient can not steepen itself). The
growth of electric field fluctuations in the far edge (away from
density gradients, see Fig.~\ref{fig9}) at large bias may be due to a new instability
driven by other free energy sources such as the flow.  It is possible
that these new fluctuations could drive the observed inward flux.
However, the measured inward flux can not be consistent with transport
modeling unless a source of plasma is present in the far edge region
($r > 28$cm) to feed the measured large inward flux. No primary
electrons are present in this region, but ionization associated with
rotation-heated bulk electrons might provide an ionization source. The
source of the apparent flux reversal in these measurements will be the
subject of future work.

\subsection{Dynamics of the confinement transition\label{dynamics}}

\begin{figure}[!htbp]
\centerline{
\includegraphics[width=3.4truein]{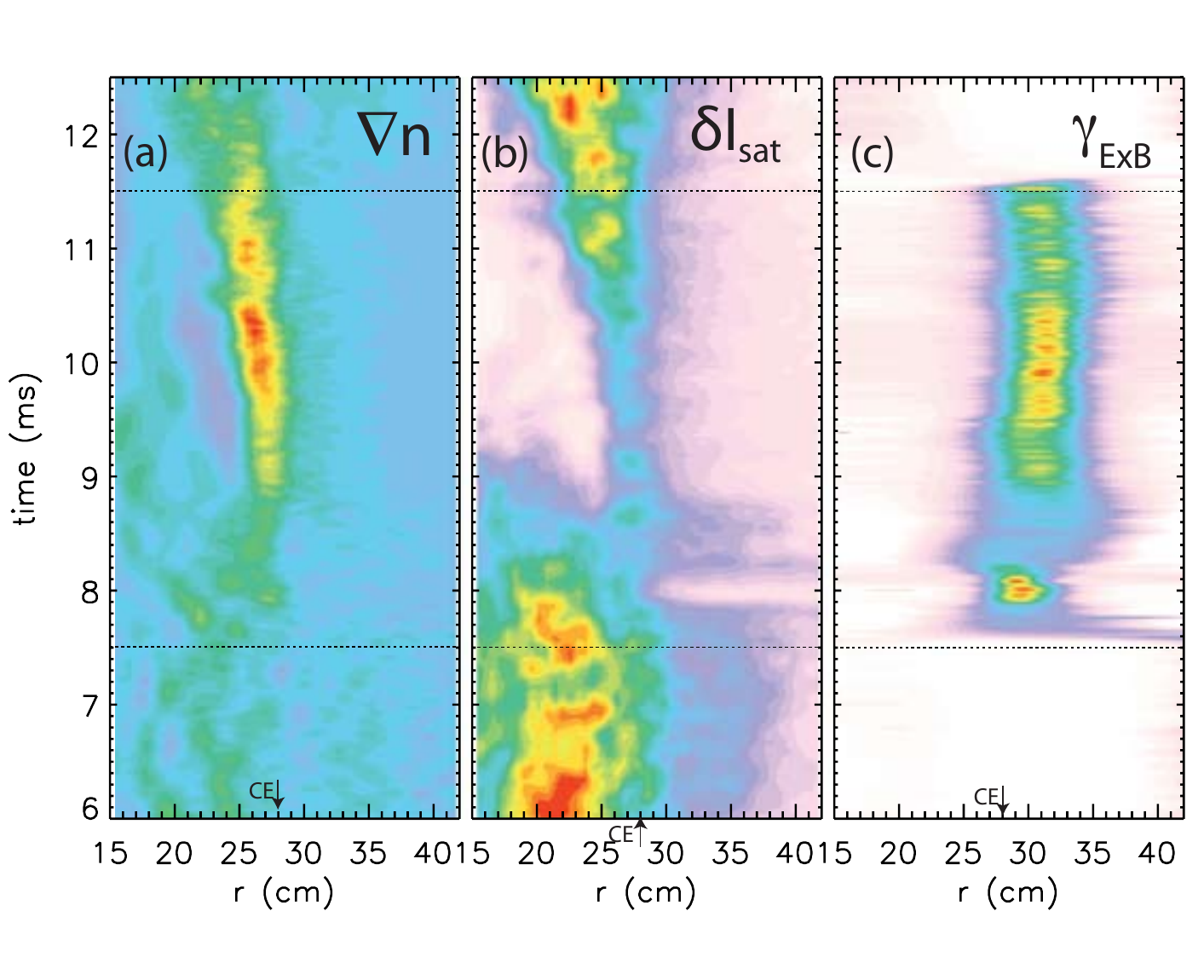}}
\caption{(color online) Contour plots of density gradient, rms
$I_{\rm sat}$ fluctuation amplitude and shearing rate versus radius and time for
125V bias.}
\label{fig12}
\end{figure}

The temporal behavior of the azimuthal flow, turbulence, and flow
profile shape during the confinement transition is interesting and
provides further support for the conclusion that the radial flow
profile is central to the transport reduction. In this section we
focus on the relation of the shear rate profile to the density
gradient profile. Figure~\ref{fig12} presents contour plots of density
gradient, rms fluctuation amplitude ($I_{\rm sat}$ fluctuations), and
shearing rate versus radius and time for a case where a confinement
transition occurs (125V bias). The 125V bias is applied from t = 7.5ms
to t = 11.5ms (both marked with dotted lines in the figure). After the
bias is turned on, the flow shear penetrates into the cathode edge,
where it peaks strongly at $t \approx 8$ms. The first effect of the
flow penetration is an immediate reduction in the fluctuation
amplitude in the edge region ($30 < r < 35$). The negative correlation
between flow shear and edge fluctuation amplitude is especially
apparent near the first strong peak in the shear ($t \approx
8$ms). The fluctuations further into the core plasma are suppressed
more slowly, but reach their lowest levels 1.5ms after the bias is
turned on. On a similar timescale, the steepening of the edge density
profile occurs and the core profile flattens. It is therefore likely
that the reduction of core fluctuations is at least in part due to
changes in the linear drive of the drift-Alfv\'en waves due to core
density profile flattening. The steepening of the edge density
gradient occurs on a timescale consistent with the transport timescale
set by end losses along the straight magnetic field
~\cite{Maggs:2007}.

\begin{figure}[!htbp]
\centerline{
\includegraphics[width=3.4truein]{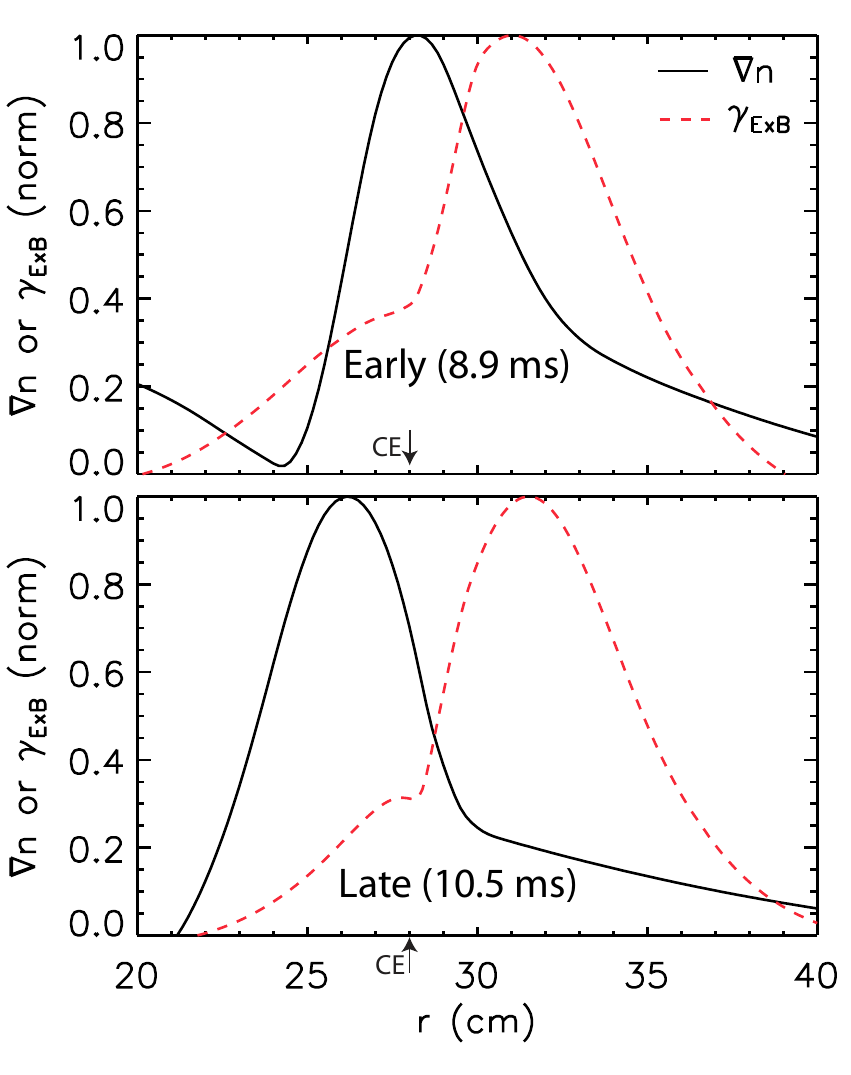}}
\caption{(color online) Profiles of density gradient and shearing
rate, derived from fits to the density and flow, early and
late in the bias pulse.}
\label{fig13}
\end{figure}

After the initial transient, the density profile steepens and the
shearing rate settles into a profile which persists in a quasi-steady
state (from $t \approx 9$ms to $t \approx 10.5$ms).  The fluctuation
  amplitude is reduced from the pre-transition value, and is localized
  to the edge gradient region. However, late in the bias pulse (near
  $t \sim 10.5$ms) the fluctuation amplitude grows up again and the
  density gradient reduces. This observation can be explained by
  considering the overlap of the shear and density gradient
  profiles. From Fig.~\ref{fig12} (b) and (c) it is apparent that late
  in the bias pulse the location of peak shear is slowly moving
  radially outward while the location of peak density gradient is
  moving radially inward. This is made clearer in Fig.~\ref{fig13}
  which shows the radial profile of density gradient and shearing
  rate, derived from fits, at two points in time. To emphasize profile
  shape and overlap, the profiles are normalized to their peak
  values. Early in the quasi-steady phase (t =8.9ms), significant
  overlap is seen between the density gradient and shearing rate
  profiles. However, later in the bias pulse (t = 10.5ms) the two
  profiles have moved apart and the amount of overlap is reduced
  substantially. Because of the reduced overlap between the density
  gradient (where linear drive for the turbulence exists) and the
  shearing rate, shear suppression of the turbulence should be less
  effective. This reduced suppression ability can explain the return
  of fluctuations and turbulent transport with the resultant reduction
  in the density gradient. Note that, since similar arguments could be
  made for the overlap of the density and flow profiles (these are the
  radial integrals of the profiles shown), it is not clear which
  profiles are determinant. However, it certainly appears that the
  degree of overlap between the profiles, whether density gradient and
  shear rate or density and flow, is crucial to the stabilization
  process.

\section{Conclusions\label{concl}}

 Biasing the central chamber wall of the LAPD, positive with respect
 to the cathode, produces a rotation layer on field lines external to
 the cathode. This rotation layer leads to a significant reduction in
 radial transport of plasma. The reduction is accompanied by changes
 in the radial temperature and density profiles of the plasma column.
 In particular the density gradient scale length is greatly reduced
 near the source edge due to the reduction in radial transport rates.
 For a fixed bias pulse length, a threshold for reduction in radial
 transport exists. The threshold voltage is apparently determined by
 radial penetration of the flow into the gradient region of the
 plasma.  Once the flow penetrates, a large flow and flow shear are
 immediately present in the gradient region and therefore a threshold
 value for the shearing rate can not be determined from this
 experiment.

 Without rotation, ion saturation current fluctuations are broadly
 distributed from the core to the plasma wall with peak amplitude very
 near, but inside, the cathode edge. As plasma rotation proceeds, the
 fluctuation profile adjusts with the changing density profile of the
 plasma. The fluctuations become concentrated in the gradient region
 of the plasma and the peak amplitude is reduced by a few tens of
 percent from the non-rotating value. The most dramatic change in the
 ion saturation fluctuations is noticed in the cross-field correlation
 function. In the non-rotating plasma, the cross-field correlation
 function is very nearly isotropic. In the rotating plasma, the radial
 correlation length is about the same as in the non-rotating
 plasma. However, the azimuthal correlation length of the rotating
 plasma is much longer than the radial correlation length (by about a
 factor of five), and the correlation function takes on a noticeable
 structure, as might be expected for a coherent cylindrical
 eigenfunction.

 The most significant change in the cross-correlation between
 azimuthal electric field fluctuations (floating potential) and
 density fluctuations (ion saturation current) occurs in the
 cross-phase in the frequency range below 10 kHz (about a tenth of the
 ion gyrofrequency).  The cross-phase changes systematically with
 increasing rotation bias.  In the unbiased, non-rotating plasma the
 cross-phase is near unity at low frequencies. As the rotation bias
 increases the cross-phase decreases toward zero and then progresses
 towards minus one for the fully rotating plasma. The measured radial
 fluxes calculated from the cross-correlation given in Eq.~\ref{eqn1}
 agree with previous modeling results for the non-rotating plasma on
 field lines outside the cathode, i.e., in the absence of fast
 tails. In the rotating plasma the cross-correlation technique
 predicts negative (inward) fluxes that are inconsistent with
 modeling.

 The change in radial transport depends upon the degree of spatial
overlap between the shear rate and density gradient profiles as
demonstrated by the behavior near threshold. The recovery of radial
transport and subsequent increase in the scale length of the density
profile, as observed late in the 125V bias case, is accompanied by a
decrease in the overlap between these two profiles. Thus the
stabilization process depends upon the spatial distribution of the
azimuthal flow.

In these experiments, azimuthal flow was driven externally, and the
interaction between the driven flow and spontaneously driven
turbulence was studied.  A critically important, unresolved issue in
confinement transitions (e.g. L- to H-mode transition) is the
mechanism behind the spontaneous generation of radial electric field
and associated sheared flows.  Many mechanisms have been proposed,
including non-ambipolar edge transport, neoclassical effects, and
turbulent processes~\cite{Itoh:1988, Itoh:1996,
  Connor:2000,Terry:2000,Tynan:2001,Diamond:2005}.  
In particular, the self-generation of flows by
turbulence (e.g. zonal flows) has recieved a great deal of attention
as a likely candidate to explain spontaneous confinement
transitions~\cite{Terry:2000,Diamond:2005}.  Direct evidence for
Reynolds-stress-driven flow has been reported in a cylindrical
device~\cite{Holland:2006} and evidence for enhanced mode-coupling prior
to the L- to H-transition has been seen in DIII-D~\cite{Moyer:2001,Tynan:2001}
(however, similar, but not exhaustive, measurements in NSTX did not
reveal enhanced mode-coupling~\cite{White:2006}).  Spontaneous
azimuthal flows are observed in LAPD prior to biasing (see
Fig.~\ref{fig2}(b)) and future work will explore the role of inverse
cascade and mode-mode coupling on generation of flows and
flow-turbulence interaction in LAPD.  Future work will also consider
the role of sheath-driven processes present in open-magnetic-field-line
configurations in flow generation~\cite{Ricci:2008}.

\section{Acknowledgements}

The authors would like to acknowledge the contributions of R.J. Taylor
and P. Pribyl in establishing the ability to induce bias-driven
rotation in LAPD.  These experiments were performed using the Basic
Plasma Science Facility at UCLA, which is supported by DOE and NSF.
TAC acknowledges support from DOE Fusion Science Center Cooperative
Agreement DE-FC02-04ER54785 and NSF Grant PHY-0547572.

%\bibliographystyle{prsty}
%\bibliography{edge_turbulence}  

\end{document}